\documentclass[11pt]{article}
\def\be{\begin{equation}}
\def\ee{\end{equation}}
\def\ba{\begin{eqnarray}}
\def\ea{\end{eqnarray}}
\def\la{\langle}
\def\ra{\rangle}
\def\a{\alpha}

\def\n{\nu}
\def\h{\hskip 1cm}

\def\lo{\longrightarrow}
\def\A1{A_{-1}}
\usepackage{epsfig}
\begin{document}
\begin{titlepage}
\vspace{4cm}
\begin{center}{\Large \bf Exact dimer ground states for a continuous family of quantum spin chains }\\
\vspace{2cm}\h  M.
Asoudeh\footnote{email:salipour@physics.iust.ac.ir},\h V. Karimipour
\footnote{Corresponding author:vahid@sharif.edu},\h L.
A. Sadrolashrafi\footnote{email:laleh@physics.sharif.edu}, \\
\vspace{1cm} $^{\dag,\ddag}$Department of Physics, Sharif University
of Technology, \\P.O. Box 11365-9161, Tehran, Iran
\end{center}
\vskip 2cm

\begin{abstract}
Using the matrix product formalism, we define a multi-parameter
family of spin models on one dimensional chains, with nearest and
next-nearest neighbor anti-ferromagnetic interaction for which exact
analytical expressions can be found for its doubly degenerate ground
states. The family of Hamiltonians which we define, depend on 5
continuous parameters and the Majumdar-Ghosh model is a particular
point in this parameter space. Like the Majumdar-Ghosh model, the
doubly degenerate ground states of our models have a very simple
structure, they are the product of entangled (dimer) states on
adjacent sites. In each of these states there is a non-zero
staggered magnetization, which vanishes when we take their
translation-invariant combination as the new ground states. At the
Majumdar-Ghosh point, these entangled states become the
spin-singlets pertaining to this model. We will also calculate in
closed form the two point correlation functions, both for finite
size of the chain and in the thermodynamic limit.

\vspace{2cm}

\end{abstract}
\hspace{.3in}
\end{titlepage}
\section{Introduction}\label{intro}
One of the basic problems of many body physics is to use analytical
and numerical tools to find the ground state of a many-body system
governed by a given Hamiltonian. Unfortunately, except for a few
exactly solvable examples, the task of finding the exact ground
state of a given Hamiltonian is notoriously difficult. Moreover in
some cases also like the anti-ferromagnetic Heisenberg chain,
although the ground state is known, its structure turns out to be
quite complicated, making the calculation of
correlation functions very difficult.  \\

In dealing with difficult problems, one way round the difficulty is
to investigate the inverse problem, that is to start from states
with pre-determined properties and find the family of Hamiltonians
for which such a state is an exact ground state. The suitable
formalism for following this path is the matrix product formalism,
which was first realized in the work of \cite{aklt}, then formalized
in finitely correlated states \cite{fcs1,fcs2}, also known as
optimal \cite{zit1,zit2} or matrix product states. This formalism
has been followed in recent years both constructing various models
of interacting spins on chains,
\cite{zit1,zit2}, ladders \cite{zitLadder,AKSLadder} and some two dimensional lattices \cite{zit2dimensional}.\\
It has also been studied to gain a better understanding of many
phenomena in correlated systems, i.e. quantum phase transitions
\cite{mpscirac}, renormalization group algorithms \cite{RG}, and
entanglement \cite{VerDC}.\\

Of particular interest to us in this paper is the so called
Majumdar-Ghosh model defined on a ring of $2N$ sites by the
following Hamiltonian
\begin{equation}\label{H0}
    H_{0} = \sum_{i=1}^{2N} 2 \sigma_i \cdot \sigma_{i+1}+
    \sigma_i\cdot \sigma_{i+2},
\end{equation}
which has been introduced and solved exactly in a series of papers
long ago \cite{mg1,mg2,mg3}, before the advent of the matrix product
formalism, and later found to have a matrix product representation
\cite{mpsgeneral}. It was shown in the original papers that the
ground state of this system is doubly degenerate, and each of these
states has a very simple structure, that is, it is a product of
singlets on adjacent sites. Let us call these ground states
$|\phi_1\ra$ and $|\phi_2\ra$, and denote a singlet state on two
adjacent sites by $S = \frac{1}{\sqrt{2}}(|01\ra-|10\ra)$, where
$|0\ra $ and $|1\ra$ denote the up and down spin in the $z$
direction, i.e. $|0\ra = |z,+\ra,\ \ |1\ra=|z,-\ra$. Then the
degenerate ground states of (\ref{H0}) are:
\begin{eqnarray}\label{phi1phi2}
|\phi_1\ra &=& S_{1,2}\ S_{3,4} \ \cdots\ \ S_{2N-1,2N}, \cr
|\phi_2\ra &=& S_{2N,1}S_{2,3}\cdots S_{2N-2,2N-1},
\end{eqnarray}
with $H_0|\phi_{1,2}\ra = -6N|\phi_{1,2}\ra$. From these two states,
one can form the new states $|\Psi_{\pm}\ra:=|\phi_1\ra\pm
|\phi_2\ra$, which are eigenstates of the single-unit translation
operator $T$ on the ring, i.e. $T|\Psi_{\pm}\ra = \pm
|\Psi_{\pm}\ra$. The correlation functions on these new translation
invariant ground states, denoted by the subscripts $\pm$, are
defined as:
$$
    \la \sigma_1^a\sigma_r^a\ra_{\pm} = \frac{\la \Psi_0^{\pm}|\sigma_1^a\sigma_r^a|\Psi_0^{\pm}\ra}{\la
    \Psi_0^{\pm}|\Psi_0^{\pm}\ra},\h a=x,\ y,\ z.
$$
These were calculated in thermodynamic limit ($N\lo \infty$) in
\cite{mg1} and were found to be
\begin{equation}\label{correlationsGM}
  \la \sigma_1^a\sigma_r^a\ra_{\pm} = -\frac{1}{2} \delta_{r,2}\h a = x, \ y, \ z.
  \end{equation}
Since then many works have been done on different aspects of this
model\cite{mg4}, including studies of spontaneous dimereization,
excitation spectrum \cite{MGExcitation} and renormalization group
study \cite{JafariLangari} of this or very closely related models
(i.e. models with anisotropy). It has also been shown \cite{kumar}
that the Majumdar-Ghosh model is the first member of a discrete
series of models, all having exact dimer ground states, where
exchange coupling decreases linearly with the separation of spins,
and vanishes beyond a certain range, $\n$ ($\n=2$ for the
Majumdar-Ghosh model).
\\

The aim of this paper is to study the Majumdar Ghosh model in the
context of matrix product formalism and in this way introduce a a
multi-parameter deformations of this model and prove that its ground
state is still doubly degenerate and dimerized. As is clear from
(\ref{H0}), the Majumdar-Ghosh model is rotationally invariant. What
we will do in this paper is that we reduce this symmetry to
rotations in the $x-y$ plane only and ask what are the Hamiltonians
on the chain with nearest and next-nearest neighbor interactions,
which are still
exactly solvable.\\

We will find a multi-parameter family of such exactly solvable
Hamiltonians and show that their ground states are doubly degenerate
and of dimer type. Both the Hamiltonian and their ground states will
reduce to those of the Majumdar-Ghosh model in a certain limit. We
will calculate all the correlation functions not only in
thermodynamic limit, but for any finite $N$. Our generalization of
this model may shed light on some aspects of the Majumdar-Ghosh
model
and provide different proofs for some of its properties.\\
Moreover, since the model is less symmetric and has 5 tunable
parameters, it may be more flexible for matching with real experimental situations. \\

The structure of this paper is as follows: first we review briefly
the matrix product formalism in section (\ref{MPS}) and then in
section (\ref{MG}) we derive the Majumdar-Ghosh model by using the
matrix product formalism. In section (\ref{DeformedMG}) we use the
same formalism to construct a continuous multi-parameter family of
models, and show that the original Majumdar-Ghosh model corresponds
to a specific point in this parameter space. In various subsections
of this section we make a detailed study of this model, derive its
Hamiltonian, determine the structure of its ground state and  prove
that the doubly degenerate ground states of this model have a simple
structure, that is, they are products of entangled spin states on
adjacent sites, in the same way that the ground states of the
Majumdar-Ghosh model are products of singlets on adjacent sites. We
also calculate all the one and the two point- functions of this
model. In particular we show that each of the degenerate ground
states has a non-zero staggered magnetization, however the
translation-invariant ground states have zero staggered
magnetization. We conclude the paper with a discussion.

\section{Matrix Product States}\label{MPS}
Let us first make a quick review of the matrix product states in a
language which we find convenient \cite{mpscirac}. For more detailed
reviews of the subject, the reader can consult more comprehensive
review articles \cite{mpsgeneral}. Consider a ring of $N$ sites,
where each site describes a $d-$level state. The Hilbert space of
each site is spanned by the basis vectors $|i\ra, \ \ i=0,\cdots ,
d-1$. A state
$$
    |\Psi\ra=\sum_{i_1,i_2,\cdots, i_N}\Psi_{i_1i_2\cdots
    i_N}|i_1i_2\cdots i_N\ra
$$
is called a matrix product state if there exist matrices $A_i,\ \
i=0,\cdots, d-1$ (of dimension $D$), such that
\begin{equation}\label{mps}
    \Psi_{i_1i_2\cdots
    i_N}=\frac{1}{\sqrt{Z}}tr(A_{i_1}A_{i_2}\cdots A_{i_N}),
\end{equation}
where $Z$ is a normalization constant. This constant is given by $$
Z=tr(E^N), $$  where
$$
E=\sum_{i=0}^{d-1} A_i^*\otimes A_i.
$$
Note that we are here considering homogeneous matrix product states
(MPS) where the matrices depend on the value of the spin at each
site and not on the site itself. More general MPS's can be defined
where the matrices depend also on the sites \cite{mpsgeneral}. It
has been shown that
such states can represent any quantum state on a chain \cite{Anders}. \\

The collection of matrices $\{A_i\}$ and $\{\mu UA_iU^{-1}\}$, where
$\mu$ is an arbitrary complex number, both lead to the same matrix
product state, the freedom in scaling with $\mu$, is due to its
cancelation with $Z$ in the denominator of (\ref{mps}).  This
freedom will be useful when we
discuss symmetries.\\

\textbf{Symmetries:} The state $|\Psi\ra$ is reflection symmetric if
there exists a matrix $\Pi$ and a scalar $\sigma$ such that
$$A_i^T=\sigma \Pi A_i\Pi^{-1} \ \ \ \ \ \ \forall \ i.$$ It is invariant under
spin flip transformation if there exists a matrix $X$ and a scalar
$\lambda$, such that $$A_{\overline i}=\lambda X A_i X^{-1}\ \ \ \ \
\ \forall \ i.$$ Finally, it is time-reversal invariant if there
exists a matrix $V$, and a scalar $\tau$ such that
$$A_i^*=\tau V A_i V^{-1}, \ \ \ \ \ \ \ \forall \ i.$$ In this article, we work
with real matrices so that the last condition is automatically
satisfied. Let us now turn to
continuous symmetries.\\

Consider a local symmetry operator $R$ acting on a site as
$R|i\ra=R_{ji}|j\ra$ where summation convention is being used. $R$
is a $d$ dimensional unitary representation of the symmetry. A
global symmetry operator ${\cal R}:=R^{\otimes N}$ will then change
this state to another matrix product state
$$
    \Psi_{i_1i_2\cdots i_N}\lo \Psi'_{i_1 i_2\cdots i_N}:=tr(A'_{i_1}A'_{i_2}\cdots
    A'_{i_N}),
$$
where
$$
    A'_i:=R_{ij}A_j.
$$
The state $|\Psi\ra$ is invariant under this symmetry if there exist
an operator $U(R)$ and a scalar $\lambda_R$ such that
\begin{equation}\label{symm}
    R_{ij}A_j=\lambda_RU^{-1}(R)A_iU(R).
\end{equation}
Repeating this transformation puts the constraint
$$
    \lambda_{R'}\lambda_R U_{R'}U_R=\lambda_{R'R}U_{R'R}.
$$
Thus $U(R)$ is a $D$ dimensional {\it projective} representation of
the symmetry $R$. In case that $R$ is a continuous symmetry with
generators $T_a$, equation (\ref{symm}), leads to
\begin{equation}\label{symmalg}
    (T_a)_{ij} A_j=[A_i,{\cal T}_a]+\mu_a A_i,
\end{equation}
where $T_a$ and ${\cal T}_a$ are the $d-$ and $D-$dimensional
representations of the Lie algebra of the symmetry, and $\mu_a$ are
numbers. In this paper we will restrict our study to the case of
$\mu_a=0$.  Equations (\ref{symm}) and (\ref{symmalg}) will be our
guiding lines in defining states with prescribed symmetries.\\

\textbf{The Hamiltonian:} Given a matrix product state, the reduced
density matrix of one site is given by
$$
    \rho_{ij}=\frac{tr((A_i^*\otimes A_j)E^{N-1})}{tr(E^N)}.
$$
This density matrix has at least $d-D^2$ zero eigenvalues. To see
this, suppose that we can find complex numbers $c_j$ such that
\begin{equation}\label{cA}
    \sum_{j=0}^{d-1}c_j A_j=0.
\end{equation}
This is a system of $D^2$ equations for $d$ unknowns which has at
least $d-D^2$ independent solutions. Any solution ${c_j}$ gives a
null eigenvector of $\rho_{ij}$. Thus the null space of the reduced
density matrix of one site comprises a subspace of at least $d-D^2$
dimension, spanned by the independent
solutions of equation (\ref{cA}).\\
The same reasoning applies to reduced density matrices of $k$
consecutive sites which are given by
$$
    \rho_{i_1\cdots i_k, j_1\cdots j_k}=\frac{tr((A_{i_1}^*\cdots A_{i_k}^*\otimes A_{j_1}\cdots A_{j_k})E^{N-k})}{tr(E^N)}.
$$
In this case the null space of the reduced density matrix of $k$
sites is spanned by the solutions of
\begin{equation}\label{cAA}
    \sum_{j_1,\cdots, j_k=0}^{d-1}c_{j_1\cdots
    j_k}A_{j_1}\cdots A_{j_k}=0.
\end{equation}
The number of independent solutions of this system of equation is
now $d^k-D^2$. Thus for the density matrix of $k$ sites to have a
null space it is sufficient (but not necessary and this is a crucial
point) that the following inequality holds
$$
    d^k\ >\ D^2.
$$
Let the null space of the reduced density matrix be spanned by the
orthogonal vectors $|e_{\a}\ra, \ \ \ (\a=1, \cdots,  s\geq
d^k-D^2)$. Then we can construct the local hamiltonian acting on $k$
consecutive sites as
$$
    h:=\sum_{\a=1}^s \lambda_{\a} |e_{\a}\ra\la e_{\a}|,
$$
where $\lambda_{\a}$'s are positive constants. These constants
together with the parameters of the vectors $|e_{\a}\ra $ inherited
from those of the original matrices $A_i$, determine the total
number of coupling constants of the Hamiltonian. If we call the
embedding of this local Hamiltonian into the sites $l$ to $l+k$ by
$h_{l,l+k}$ then the full Hamiltonian on the chain is written as
$$
    H=\sum_{l=1}^N h_{l,l+k}.
$$
The state $|\Psi\ra$ is then a ground state of this hamiltonian with
vanishing energy. The reason is as follows:
$$
\la \Psi|H|\Psi\ra=tr(H|\Psi\ra\la\Psi|)=\sum_{l=1}^N
tr(h_{l,l+k}{\rho}_{l,l+k})=0,
$$
where $\rho_{l,k+l}$ is the reduced density matrix of sites $l$ to
$l+k$ and in the last line we have used the fact that $h$ is
constructed from the null eigenvectors of $\rho$ for $k$ consecutive
sites. Given that $H$ is a positive operator, this proves the
assertion. \\

\textbf{Correlation functions:} Let $O$ be any local operator acting
on a single site. Then we can obtain the one point function on site
$k$ of the chain $\la \Psi|O(k)|\Psi\ra $ as follows:
\begin{equation}\label{1pointMPS}
    \la \Psi|O(k)|\Psi\ra = \frac{tr(E^{k-1}E_O E^{N-k})}{tr(E^N)},
\end{equation}
where
$$
E_O:=\sum_{i,j=0}^{d-1}\la i|O|j\ra A_i^*\otimes A_j.
$$
In the thermodynamic limit $N\lo \infty$, equation (\ref{1pointMPS})
gives
$$
    \la \Psi|O|\Psi\ra = \frac{\la \lambda_{max}|E_O|\lambda_{max}\ra}{\lambda_{max}}
$$
 where we have used the translation invariance of the model and
$\lambda_{max}$ is the eigenvalue of $E$ with the largest absolute
value, for which the matrix element is non-vanishing, and
$|\lambda_{max}\ra$ and $\la \lambda_{max}|$ are the right and left
eigenvectors corresponding to this eigenvalue, normalized such that
$\la \lambda_{max}|\lambda_{max}\ra=1$. Here we are
assuming that the largest eigenvalue of $E$ is non-degenerate.\\

The n-point functions can be obtained in a similar way. For example,
the two-point function $\la \Psi|O(k)O(l)|\Psi\ra$ can be obtained
as
\begin{equation}\label{2pointMPS}
\la \Psi|O(k)O(l)|\Psi\ra = \frac{tr(E_O(k)E_O(l)E^N)}{tr(E^N)}
\end{equation}
where $E_O(k):=E^{k-1}E_OE^{-k}$. Note that this is a formal
notation which allows us to write the n-point functions in a uniform
way, it does not require that $E$ is an invertible matrix. In the
thermodynamic limit the two point function turns out to be
$$
\la \Psi|O(1)O(r)|\Psi\ra = \frac{1}{\lambda_{max}^{r}} {\sum_i
\lambda_i^{r-2} \la\lambda_{max}|E_{O}|\lambda_{i}\ra\la
\lambda_i|E_{O}|\lambda_{max}\ra}.
$$

\section{The Majumdar-Ghosh model}\label{MG}
We take the size of the ring an even number $2N$. Let us also choose
$d=2$ (spin $1/2$ particles), $k=3$ (interactions up to next-nearest
neighbors) and work with $D=3$ dimensional matrices. Note that the
condition $d^k> D^2$ is not satisfied here, but in view of the fact
that this is a sufficient and not necessary condition, this will not
be a source of concern. The basic symmetry that we demand is the
rotational symmetry around the $z$ axis, which according to
(\ref{symmalg}) demands that there be a matrix $S_z$ such that
$$
    [A_0,S_z]=\frac{1}{2}A_0,\h [A_1,S_z]=-\frac{1}{2}A_1.
$$
Working in a basis where $S_z=diagonal (-1/2, 0, 1/2)$, these
equations immediately show that $A_0$ and $A_1$ should be of the
following form:
$$
    A_0:=\left(\begin{array}{ccc} 0 & p & 0 \\ 0 & 0 & q \\ 0 & 0 & 0
    \end{array}\right)\ \ \ \ \ \  , \ \ \ \ \  A_1:=\left(\begin{array}{ccc} 0 & 0 & 0 \\ h & 0 & 0 \\ 0 & g & 0
    \end{array}\right).
$$
By a similarity transformation $A_i\lo SA_i S^{-1}$ where
$S=diagonal (1, p, pq)$, the parameters of the matrix $A_0$ are both
set equal to $1$, (the parameters of $A_1$ change which we again
denote them by $g$ and $h$.)  At this stage no similarity
transformation or re-scaling, can reduce the number of parameters,
however we note that the matrices $A_0$ and $A_1$ act as raising and
lowering operators on the 3 dimensional space they act on, ($i.e.
A_0|i\ra=|i+1\ra,\ A_1|i\ra\propto |i-1\ra$) and hence the trace of
any string of operators $A_{i_1}A_{i_2}\cdots A_{i_{2N}}$ is
non-zero only if the number of $A_0$ and $A_1$ operators are equal
in such a string. This implies first that any such string should be
of even length (that is why we have taken a ring of even length),
and second that a factor of $h^N$ will be removed from the numerator
and denominator of (\ref{mps}) due to the normalization of the
state.

Thus the final matrices which will make a matrix product state with
$S_z$ symmetry, will be of the form:
\begin{equation}\label{MatricesSzScaling}
    A_0:=\left(\begin{array}{ccc} 0 & 1 & 0 \\ 0 & 0 & 1 \\ 0 & 0 & 0
    \end{array}\right)\ \ \ \ \ \  , \ \ \ \ \  A_1:=\left(\begin{array}{ccc} 0 & 0 & 0 \\ 1 & 0 & 0 \\ 0 & g & 0
    \end{array}\right).
\end{equation}
This is all we can go with $S_z$ symmetry and we will in fact make a
detailed study of such models in the next section.\\

What happens if we impose the discrete spin-flip symmetry on the
state? This demands that there be a matrix $X$ and a scalar
$\lambda$ such that
$$
    A_0 = \lambda X A_1 X^{-1},\h A_1 = \lambda X A_0 X^{-1}.
$$
A short calculation shows that the matrix $X$ which satisfies the
above equation is equal to
$$
    X = \left(\begin{array}{ccc} 0 & 0 & 1 \\ 0 & \lambda & 0 \\ \lambda^2 & 0 & 0
    \end{array}\right),
$$
where $\lambda^2=g=1$ . Thus spin-flip symmetry requires that the
parameter $g$ be a discrete parameter which we rename as
$g=\epsilon$, where $\epsilon=\pm 1$.\\

It is now found that the parity symmetry imposes no new condition on
the parameters, since with the following matrix
$$
        \Pi = \left(\begin{array}{ccc} 0 & 0 & 1 \\ 0 & \sqrt{\epsilon} & 0 \\ \epsilon & 0 & 0
    \end{array}\right),
$$
one can verify that $ A_i^T = \sqrt{\epsilon} \Pi A_i \Pi^{-1}. $
Therefore the final form of the matrices leading to a model with a
continuous $S_z$ symmetry plus two discrete parity and spin flip
symmetries will be of the form:
\begin{equation}\label{MatricesSzSpinFlipParity}
    A_0:=\left(\begin{array}{ccc} 0 & 1 & 0 \\ 0 & 0 & 1 \\ 0 & 0 & 0
    \end{array}\right)\ \ \ \ \ \  , \ \ \ \ \  A_1:=\left(\begin{array}{ccc} 0 & 0 & 0 \\ 1 & 0 & 0 \\ 0 & \epsilon  & 0
    \end{array}\right).
\end{equation}

How is the matrix product state (\ref{mps}) with the above matrices
is related to the states $|\phi_{1,2}\ra$? We will show in the next
section, when we study the generalization of this model, that the
matrix product state is equal to $|\phi_1\ra + |\phi_2\ra$, that is
it is the linear superposition of the ground states which is
invariant under translation (with eigenvalue 1). In fact it has been
shown that both $|\phi_1\ra$ and $|\phi_2\ra$ are ground states and
so translation invariant combinations of them, namely
$|\Psi_{\pm}\ra:=|\phi_1\ra \pm |\phi_2\ra$, are also ground states.
The point is that only $|\Psi_+\ra$ has a matrix product
representation. It sometimes happens that from a single matrix
product state, one can derive the degenerate grounds states of a
model not all of which have matrix
product representations. The Majumdar-Ghosh model is one such case. \\

Let us also find the Hamiltonian. With the matrices
(\ref{MatricesSzSpinFlipParity}) we should solve the equations
$\sum_{i,j,k=0}^1 c_{ijk}A_iA_jA_k=0$. It is readily found that the
solution space of this system is spanned by the following four
vectors:
\begin{eqnarray}\nonumber
  |e_1\ra  &=& |000\ra \cr
  |e_2\ra &=& |111\ra \cr
  |e_3\ra &=& |001\ra - \epsilon|010\ra +|100\ra \cr
  |e_4\ra &=& |011\ra - \epsilon|101\ra +|110\ra.
\end{eqnarray}
These vectors are parity invariant (invariant under left-right
reflection) and are mapped to each other under spin-flip
transformation ($|0\ra\rightarrow |1\ra,\ |1\ra\rightarrow |0\ra$).
A Hamiltonian with all three symmetries is constructed as:
$$
  h=J(|e_1\ra\la e_1|+|e_2\ra\la e_2|)+K(|e_3\ra\la e_3|+|e_4\ra\la
    e_4|),
$$
which when expanded in terms of the spin operators, and embedded
into the chain, makes the total Hamiltonian equal to the following:
$$
    H = \sum_{i=1}^N -K\epsilon \sigma_{i}\cdot \sigma_{i+1} + \frac{K}{2}\sigma_{i}\cdot \sigma_{i+2} +
\frac{J+(2\epsilon-1)K}{2}\sigma_{i,z}\sigma_{i+1,z}+\frac{J-3K}{4}
\sigma_{i,z}\sigma_{i+2,z}.
$$
For $\epsilon=1$ this reduces to
$$
    H_{1} =\sum_{i=1}^{2N} -K \sigma_{i}\cdot \sigma_{i+1} +\frac{K}{2} \sigma_{i}\cdot \sigma_{i+2} +
\frac{J+K}{2}\sigma_{i,z}\sigma_{i+1,z}+\frac{J-3K}{4}
\sigma_{i,z}\sigma_{i+2,z},
$$
and for $\epsilon=-1$ reduces to
$$
    H_{-1} = \sum_{i=1}^{2N} K \sigma_{i}\cdot \sigma_{i+1} +\frac{K}{2} \sigma_{i}\cdot \sigma_{i+2} +
\frac{J-3K}{4}(2\sigma_{i,z}\sigma_{i+1,z}+\sigma_{i,z}\sigma_{i+2,z}).
$$
It is the Hamiltonian $H_{-1}$ which turns into the already familiar
Majumdar-Ghosh model (\ref{H0}), by taking $J=3K$. For general
values of $J$ and $K$ it describes the anisotropic Majumdar-Ghosh
model, where the anisotropy parameter $\frac{J_z}{J_x=J_y}$ is the
same for
both the nearest and next-nearest neighbors.\\

We do not know whether this is already well-known that this model is
also exactly solvable. In fact one can check directly that the
eigenstates (\ref{phi1phi2}) are also eigenstates of the operator
$$H_z:=\sum_{i=1}^{2N} 2
\sigma^z_{i}\sigma^z_{i+1}+\sigma^z_{i}\sigma^z_{i+2},$$ with
eigenvalue $2N$. To see this one needs to invoke the following
relations which hold true for any two consecutive singlets $S_{12}$
and $S_{34}$:
\begin{eqnarray}\nonumber
\sigma_1^z\sigma_{2}^zS_{_{1,2}}S_{_{3,4}}&=&-S_{_{1,2}}S_{_{3,4}}\cr
\sigma_{1}^z\sigma_{3}^zS_{_{1,2}}S_{_{3,4}}&=&U_{_{1,2}}U_{_{3,4}}\cr
\sigma_{2}^z\sigma_{3}^zS_{_{1,2}}S_{_{3,4}}&=&-U_{_{1,2}}U_{_{3,4}},
\end{eqnarray}
where $U = \frac{1}{\sqrt{2}}(|01\ra+|10\ra)$.

 Using these relations one sees that in $H_z|\phi_{1,2}\ra$
all the terms containing the $U$ terms cancel out and one is left
with $H_z|\phi_{1,2}\ra=-2N|\phi_{1,2}\ra$.

Putting all these together we find
$$
    H_{-1}|\phi_{1,2}\ra = \frac{-J-3K}{2}N|\phi_{1,2}\ra.
$$

What is the Hamiltonian $H_1$? Does it represent any new model? The
answer is negative. To see this we note that by rotating the
even-numbered spins around the $z$ axis by an angle of $\pi$, under
which $\sigma^{x,y}_{2i}\lo -\sigma^{x,y}_{2i}$ and
$\sigma^z_{2i}\lo \sigma^z_{2i}$, while keeping the odd-numbered
spins intact, the Hamiltonian $H_1$ is mapped onto $H_{-1}$.\\
This transformation will map the states $|\phi_{1,2}\ra$ onto the
states $|\chi_{1,2}\ra$ where

\begin{eqnarray}\nonumber |\chi_1\ra &=& U_{1,2}U_{3,4}\ \cdots \ U_{2N-1,2N}, \cr
|\chi_2\ra &=& U_{2N,1}U_{2,3}\cdots U_{2N-2,2N-1}. \end{eqnarray}

These states are the doubly degenerate ground states of $H_1$.

\section{Deformation of the Majumdar-Ghosh model}\label{DeformedMG}
In the last section we have shown how the Majumdar-Ghosh Hamiltonian
and its ground states are derived in the matrix product formalism.
Let us now relax the condition of parity symmetry, which
automatically removes also the spin flip symmetry. Then the matrices
will be of the form (\ref{MatricesSzScaling}), which still have the
continuous symmetry of rotation around the $z$ axis. It is now
easily verified that the solution space of the system of equations
(\ref{cAA}) with $k=3$ is spanned by the following four vectors:
\begin{eqnarray}\nonumber
  |e_1\ra  &=& |000\ra \cr
  |e_2\ra &=& |111\ra \cr
  |e_3\ra &=& |001\ra - g|010\ra +g^2|100\ra \cr
  |e_4\ra &=& |011\ra - g|101\ra +g^2|110\ra.
\end{eqnarray}
 The most general local Hamiltonian having the matrix product state
 as its ground state is now given by
$$
    h=J_1|e_1\ra \la e_1| + J_2|e_2\ra \la e_2|+J_3|e_3\ra \la
    e_3|+J_4|e_4\ra\la e_4|.
$$
Rewriting this in terms of spin operators we find after some rather
lengthy calculations the following form of the final Hamiltonian
\begin{eqnarray}\label{HDeformed}
H =\frac{1}{8}\sum_{i=1}^N &K_1&\  {\sigma} _i \cdot {\sigma}_{i+1}
+K_2\ {\sigma}_i\cdot {\sigma}_{i+2} + K_3 \
\sigma^z_{i}\sigma^z_{i+1} + K_4  \sigma^z_{i}\sigma^z_{i+2} + K_5
\sigma^z_{i} +K_6 \sigma^z_{i}\sigma^z_{i+1}\sigma^z_{i+2} \cr +
&K_7& \sigma^z_{i}{\sigma}_{i+1}{\sigma}_{i+2}+ K_8 {\sigma}_i\cdot
{\sigma}_{i+1}\sigma^z_{i+2}+K_9\ {\sigma}_{i}\cdot
\sigma^z_{i+1}{\sigma}_{i+2},
\end{eqnarray}
where the couplings are
\begin{eqnarray}\nonumber
K_1 &=& -2g(1+g^2)(J_3+J_4) \cr K_2 &=& 2g^2(J_3+J_4) \cr K_3 &=&
2(J_1+J_2)+2(J_3+J_4)(g-g^2+g^3)\cr K_4 &=&
J_1+J_2-(J_3+J_4)(1+g^2+g^4)\cr K_5 &=&
3(J_1-J_2)+(J_3-J_4)(1+g^2+g^4) \cr K_6 &=&
J_1-J_2-(J_3-J_4)(1-g+g^2)^2 \cr K_7 &=& -2g J_3+2g^3 J_4 \cr K_8
&=& -2g^3 J_3+2g J_4\cr K_9 &=& 2g^2(J_3-J_4).
\end{eqnarray}
This is a 5-parameter family of models with nearest neighbor and
next-nearest neighbor interactions, whose ground state can be
written down exactly. We will show that with the couplings as above,
this model has a doubly degenerate ground state, each of which has a
simple product structure of dimers, like the one given in (\ref{phi1phi2}).\\

Note that the Hamiltonian (\ref{HDeformed}) will reduce to $H_{-1}$
when we restrict the parameters as follows
$$g=-1, J_1=J_2=J, J_3=J_4=K$$ and to (\ref{H0}) when we further set $J=3K$.\\

\subsection{Structure of the matrix product state}

The matrix product state (\ref{MatricesSzScaling}) has a simple
structure.  We prove the following theorem:\\

\textbf{Theorem:} The matrix product state (\ref{mps}) constructed
from the matrices (\ref{MatricesSzScaling}) which is a ground state
of (\ref{HDeformed}) has the following structure. Define a state

$$
   S(g):=g|01\ra+|10\ra,
$$
\begin{eqnarray}\nonumber
  |\phi_1(g)\ra &=& S_{_{1,2}}(g) S_{_{3,4}}(g)S_{_{5,6}}(g)\cdots S_{_{2N-1,2N}}(g) \\
  |\phi_2(g)\ra &=& S_{_{2N,1}}(g) S_{_{2,3}}(g)S_{_{4,5}}(g)\cdots
  S_{_{2N-2,2N-1}}(g),
  \end{eqnarray}
that is $|\phi_2(g)\ra = T|\phi_1(g)\ra$, where $T$ is the
translation operator along the chain by one lattice unit. Then we
have
$$
|\Psi(g)\ra = |\phi_1(g)\ra + |\phi_2(g)\ra.
$$

\textbf{Proof:} Consider the matrix product state
$|\Psi(g)\ra=\sum\Psi_{i_1i_2\cdots i_{2N}}|i_1i_2\cdots i_{2N}\ra$.
The amplitudes
$\Psi_{i_1i_2\cdots i_{2N}}$ are non-vanishing only when\\

i) the number of $0$ indices and $1$ indices are equal (in view of
the fact that the matrices $A_0$ and $A_1$ are raising and lowering
operators), ii) no three consecutive indices are $0$ or $1$, (since
$A_0^3=A_1^3=0$), and iii) no substrings of the types $\cdots
11011\cdots$ or $\cdots 00100\cdots$ appear in the indices, (since
$A_1^2A_0A_1^2=A_0^2A_1A_0^2=0$.)\\

Let us define the compact notations ${\bf{0}}:=01$ and ${\bf
{1}}:=10$. In view of the above three properties, we can divide the
set of states $|i_1,i_2,\cdots i_{2N}\ra$ into two types, type 1,
which are states of the form $|{\bf 0}{\bf 0}{\bf 1}{\bf 1}{\bf 0}
\ra=|0101101001\ra$, that is, strings of ${\bf 0}$'s and ${\bf 1}$,s
in arbitrary order, and type 2, which are transformed to the above
type by a one-unit cyclic translation like $|1010110100\ra$. There
is a one to one correspondence between these two types of states and
hence the state $|\Psi\ra$ can be divided in two parts, depending on
the type of basis states \ $  |\Psi(g)\ra =
|\phi_1(g)\ra+|\phi_2(g)\ra \ $, where
$$\phi_1(g)=\sum_{\a_1, \a_2, \cdots , \a_N={\bf 0},{\bf
1}}\phi_{\a_1 \a_2 \cdots \a_N}|\a_1 \a_2 \cdots \a_N\ra$$ and
$|\phi_2(g)\ra=T|\phi_1(g)\ra$. We now note that
\begin{equation}\label{A0A1}
   A_{{\bf 0}}\equiv A_0A_1=\left(\begin{array}{ccc} 1 &&\\ & g& \\
    &&\end{array}\right)\h A_{{\bf 1}}\equiv A_1A_0 =\left(\begin{array}{ccc}  &&\\ & 1& \\
    &&g\end{array}\right).
\end{equation}
This means that $A_{\bf 0}$ and $A_{{\bf 1}}$ commute with each
other, which immediately implies that $|\phi_1(g)\ra$ is a product
state, i.e.
$$
    |\phi_1(g)\ra = \sum_{m=0}^N tr(A_{{\bf 0}}^mA_{{\bf
    1}}^{N-m})|{\bf 0}^m{\bf 1}^{N-m}\ra =\sum_{m=0}^N g^m|{\bf 0}^m{\bf 1}^{N-m}\ra
    =(g|{\bf 0}\ra+|{\bf 1}\ra)^{\otimes N},
$$
where $|{\bf 0}^m{\bf 1}^{N-m}\ra$ denotes a uniform superposition
of states each of which has $m$ ${\bf 0}'s$ and $N-m$ ${\bf 1}$'s,
in different positions. This completes the proof.\\

Are the two individual states $|\phi_1(g)\ra$ and $|\phi_2(g)\ra$
the ground states of $H$? There are supports for thinking so. First
we have checked this for low values of system size $N$ and second we
know that as $g\lo -1$, these states approach the states of
Majumdar-Ghosh (\ref{phi1phi2}) which are known to be ground states
of the Hamiltonian $H_{-1}$. Therefore we can make two linear
combinations of these ground states, namely $|\Psi_+(g)\ra :=
|\phi_1(g)\ra+|\phi_2(g)\ra$ which has a matrix product
representation and $|\Psi_-(g)\ra := |\phi_1(g)\ra-|\phi_2(g)\ra$
which does not have such a representation. In the next subsection we
calculate the correlation functions on these translation invariant
states and they again approach the ones calculated in \cite{mg1} in
the limit $g=-1$ and $N\lo \infty$. For our future reference we need
the following easily verified properties:
$$
    \la \phi_1(g)|\phi_1(g)\ra = \la \phi_2(g)|\phi_2(g)\ra = (1+g^2)^{N}\h  \la \phi_1(g)|\phi_2(g)\ra = \la \phi_2(g)|\phi_1(g)\ra =
    2g^{N}.
$$

\subsection{Correlation functions}
In the Majumdar-Ghosh model both the ground states (\ref{phi1phi2})
have anti-ferromagnetic order and $\la \sigma_i^{x,y,z}\ra=0$. This
is clear from the fact that the states (\ref{phi1phi2}) are product
of singlets on adjacent sites. In our model, the states are no
longer singlets and while there is no magnetic order present, the
staggered magnetization is non-vanishing. Consider the states
$|\phi_1(g)\ra$, and $|\phi_2(g)\ra$. In view of their simple
product structure, it is readily seen that

\begin{equation}\label{SzPhi1}
    \la\phi_1(g)| \sigma_1^z|\phi_1(g)\ra= -\la\phi_1(g)| \sigma_2^z|\phi_1(g)\ra=\la
    S_{12}|\sigma_1^z|S_{12}\ra = \frac{1-g^2}{1+g^2},
\end{equation}
and
\begin{equation}\label{SzPhi2}
    \la\phi_2(g)| \sigma_1^z|\phi_2(g)\ra=-\la\phi_2(g)|
    \sigma_2^z|\phi_2(g)\ra=\la
    S_{2N,1}|\sigma_1^z|S_{2N,1}\ra = \frac{g^2-1}{1+g^2}.
\end{equation}
This means that the value of staggered magnetization defined as
$$(m_s)_{\phi_i}:=\frac{1}{2N} \sum_{i=1}^N \la \phi_i|\sigma_{z,2i-1}-\sigma_{z,2i}
|\phi_i\ra$$ is non-zero in these two states, i.e.

$$(m_s)_{\phi_1}=\frac{1-g^2}{1+g^2},\h (m_s)_{\phi_2}=\frac{g^2-1}{1+g^2}.$$

However there is no staggered magnetization in the translation
invariant states $|\Psi_{\pm}\ra$. To see this we can  calculate
$\sigma^z_1$ for the state $|\Psi_+\ra=|\phi_1(g)\ra+|\phi_2(g)\ra$,
either by using the matrix product formula (\ref{1pointMPS}) or by a
direct calculation in which we use (\ref{SzPhi1}), (\ref{SzPhi2})
and

\begin{equation}\label{CorrelationsPsi}
    \la \Psi_{\pm}| O |\Psi_{\pm}\ra =  \frac{\la \phi_1|O|\phi_1\ra + \la \phi_2|O|\phi_2\ra \pm 2\la \phi_1|O|\phi_2\ra}{2\la
    \phi_1|\phi_1\ra\pm 2 \la \phi_1|\phi_2\ra},
\end{equation}
in which $O$ is any operator and we have abbreviated $|\phi_i(g)\ra$
to $|\phi_i\ra$ and used the fact that $\la \phi_1|\phi_1\ra=\la
\phi_2|\phi_2\ra$ to simplify terms in the denominator. For this
direct calculation, the only new quantity which is needed is $\la
\phi_1(g)|\sigma_1^z|\phi_2(g)\ra$. We know that all the states in
the expansion of $|\phi_1(g)\ra$ and $|\phi_2(g)\ra$ are different
from each other except the states $|{\bf 0}{\bf 1}{\bf 0}\cdots {\bf
1}\ra$ and $|{\bf 1}{\bf 0}{\bf 1}\cdots {\bf 0}\ra$ which are
common to both and transformed into each other by the action of the
translation operator $T$. So we find
\begin{equation}\label{SzPhi1Phi2}
    \la \phi_1(g)|\sigma_1^z|\phi_2(g)\ra=(g^n\la {\bf 0}{\bf 1}\cdots {\bf
1}| + \la {\bf 1}{\bf 0}\cdots {\bf 0}|) |\sigma_1^z|(g^n| {\bf 1}
{\bf 0}\cdots {\bf 0}\ra+|{\bf 0}{\bf 1}\cdots {\bf 1}\ra)
 =0,\end{equation}
implying that there is no staggered magnetization in
$|\Psi_{\pm}\ra$.

The two point functions can also be calculated along similar lines.
We point out only the general method. The product nature of
$|\phi_1\ra$ and $|\phi_2\ra$, formula (\ref{CorrelationsPsi}) and
the reasoning used above for the calculation of the cross term $\la
\phi_1(g)|\sigma_1^z\sigma_r^z|\phi_2(g)\ra$ facilitate calculations
of $\la  \sigma_1^z\sigma_r^z\ra_{\pm}$. However the cross term $\la
\phi_1(g)|\sigma_1^x\sigma_r^x|\phi_2(g)\ra$ is hard to calculate in
this way. Instead we calculate $\la \sigma_1^x\sigma_r^x\ra_+$ (i.e.
for the matrix product state $|\Psi_+\ra$) from (\ref{2pointMPS})
and infer the cross term from its result. We then use this cross
term for determining the correlation $\la
\sigma_1^x\sigma_r^x\ra_-$. The reader can verify these results by
direct calculation for chains of small size. The interesting point
is that all the correlation functions can be expressed in terms of a
variable $u:=g+\frac{1}{g}$ which is invariant under the interchange
of $g\lo \frac{1}{g}$. The results are:
$$
    \la \sigma^z_{1}\sigma^z_2\ra_{\pm} =
    \frac{-2\ \pm u^{N}(2u^{-2}-1)}{2\ \pm u^N} , \h
    \la \sigma^z_{1}\sigma^z_{r>2}\ra =
    (-1)^{r}\frac{-2 \pm u^{N}(4u^{-2}-1)}{2 \pm \ u^N},
$$

and
$$
    \la \sigma^n_{1}\sigma^n_2\ra_{\pm} =
    \frac{u\ \pm u^{N-1}}{2\ \pm u^N}, \h
    \la \sigma^n_{1}\sigma^n_{r>2}\ra =
    \frac{2\delta_{r,odd} + \delta_{r,even} u}{2 \pm \ u^N}.
$$
It can be verified that these correlation functions reduce to
(\ref{correlationsGM}) at the point $g=-1$ ($u=-2$), after we take
the thermodynamic limit ($N\lo \infty$).

\section{Discussion}
We have constructed a 5-parameter family of spin-Hamiltonians with
nearest and next nearest neighbor interactions and have found their
exact ground states. The ground states, depend on only one of these
parameters, are doubly degenerate and have a very simple structure.
They are product of entangled states (dimers) on adjacent spins. We
have also calculated the one and the two point functions in closed
analytic form both for finite size of the chain and in the
thermodynamic limit.  The dimerized ground states show staggered
magnetization, but the translation-invariant combinations of these
states have no staggered magnetization.

\section{Acknowledgement} We would like to thank A. Langari and R. Jafari, for very valuable discussions and the members of the
Quantum information group of Sharif University, specially S. Alipour
and L. Memarzadeh for instructive comments.

{}

\end{document}